\documentstyle[twoside, epsf]{article}

\input ibvs2.sty

\begin{document}

\IBVShead{}{}

\IBVStitle{Photometric and spectroscopic variations}\\
\IBVStitle {of the Be star HD\,112999}

\IBVSauth{\noindent M. A. Corti$^{1,2}$, R. C. Gamen$^{2,3}$, Y. J. Aidelman$^{2,3}$, G. A. Ferrero$^{2,3}$ \& W. A. Weidmann$^{4}$}

\IBVSinst{Instituto Argentino de Radioastronom\'{\i}a (CCT-La Plata, CONICET),Villa Elisa, Argentina}
\IBVSinst{Facultad de Ciencias Astron\'omicas y Geof\'{\i}sicas, Universidad Nacional de La Plata, Argentina \\
e-mail: {mariela@fcaglp.unlp.edu.ar}}
\IBVSinst{Instituto de Astrof\'{\i}sica de La Plata (CCT-La Plata, CONICET), Argentina}
\IBVSinst{Observatorio Astron\'omico C\'ordoba, Universidad Nacional de C\'ordoba (CCT-C\'ordoba, CONICET), Argentina}

\SIMBADobj{HD\,112999}
\IBVStyp{ Be star }
\IBVSkey{Stars: emission-line, Be - Stars: individual: HD\,112999 - Stars: variables: general}
\IBVSabs{Be objects are stars of B spectral type showing lines of the Balmer series in emission. The presence of these lines is attributed to the existence of an extended envelope, disk type, around them. Some stars are observed in both the Be and normal B-type spectroscopic states and they are known as transient Be stars.}
\IBVSabs{In this paper we show the analysis carried out on a new possible transient Be star, labelled HD\,112999, using spectroscopic optical observations  and  photometric data.}

\begintext

\section{Introduction}

Be stars were defined as `A non-supergiant B-type star whose spectrum has 
or had at time, one or more Balmer lines in emision"  by Jaschek et al. (1981).
This class of stars have circumstellar disks that originate hydrogen, helium, and/or metal lines in emission.
The main causes that are thought to generate these disks are: rapid rotation and/or binarity,
combined with non-radial pulsations  (Huat et al., 2009; Neiner et al., 2012). 
These phenomena can take place as a stage in the evolutionary life of hot B-type stars.

One of the most intriguing properties of Be stars is their long-term variability, which 
seems to be related to disk growth or loss events (Hubert \& Floquet, 1998).
These events have spectroscopic and photometric effects, i.e. changes from emission to 
absorption line (and vice versa), and also flux and color variations.
Given the great variety of observed events, many open questions on the Be stars remain
(see the review by Porter \& Revinius, 2003, and references therein), which raises the 
importance of studying these objects.
Moreover as the same authors state ``Be stars are in a unique position to make contributions to several 
important branches of stellar physics, e.g., asymmetric mass$-$loss processes, stellar angular 
momentum distribution evolution,  asteroseismology, and magnetic field evolution''.

The primary goal of this work is to show the analysis carried out on a new possible transient Be star, 
labelled HD\,112999 ($\alpha_{J2000}=$13:01:35, $\delta_{J2000}= -$60:40:16). This star was observed 
by one of us (MAC) while trying to identify possible members of the Centaurus OB1 (CenOB1) 
stellar association (Corti \& Orellana, A\&A in press). 
The spectral classification of this star presents some discrepancies in the available bibliography.
It was first reported as Be star by Bidelman \& MacConnell (1973). 
Other authors reported different spectral types, i.e. 
Cannon \& Pickering (B8; 1920), Houk \& Cowley (B6~III (N); 1975), and Garrison et al. (B6~V; 1977).
 Nowadays, HD\,112999 is listed in the BeSS 
catalogue\footnote{http://basebe.obspm.fr} as B6 III ne (Neiner et al., 2011).
As our optical spectrum, taken on 2009, was very different from those classified before, the star deserved
careful consideration. We searched spectral and photometric databases and found that HD\,112999
is a variable Be star which seems to have experimented disk $-$ loss and growth $-$ events over the last decades. 
In the following sections, we describe these facts.

\section{Observational Data}
\subsection{Spectroscopic data}

The observational material belongs to different-programs runs obtained at the 
Complejo Astron\'omico El Leoncito (CASLEO\footnote{Operated 
under agreement between CONICET and the Universities of La Plata, C\'ordoba and San Juan, Argentina.}), 
Argentina and La Silla Observatory, Chile
 In these runs were used different instrumental configurations which are detailed in Table~1. 

For CASLEO observations, comparison arc images were observed
at the same sky position as the science targets as the stellar images 
immediately after or before the stellar exposures,
being Th$-$Ar for the echelle mode of the REOSC spectrograph, Cu$-$Ar for the simple mode, and
He$-$Ne$-$Ar if the B\&C was used. Moreover, bias frames, standard stars of radial$-$velocity, 
and spectral$-$type (the last during the run B only) were observed every night. 
We have also taken the spectrum of the standard of rotational$-$velocity, $\tau$ Sco 
(Slettebak et al., 1975), during run C.
These spectra were processed and analysed with 
IRAF\footnote{IRAF is distributed by the National Optical Astronomy Observatories, which are operated by
the Association of Universities for Research in Astronomy, Inc., under cooperative
agreement with the National Science Foundation.} routines.

The spectrum retrieved from the ESO database, 
was acquired with the FEROS spectrograph and processed with the MIDAS pipeline. 
The radial velocities were obtained by simple crosscorrelation of the echelle 
orders of the object and of the simultaneous calibration fiber with respect to
a corresponding reference spectrum.

{\small
\begin{center}
\begin{table*}
\caption{Instrumental configurations used}
\begin{tabular}{cllccccccc}
\hline
\hline
run &Observat. &Epoch(s)& HJD             &Telesc. & Spectrograph& Rec. disp. & $\Delta\lambda$& exp.time& S/N\\
      &                     &                 & 2400000+ &         [m]     &                         & [\AA~px$^{-1}$] & [\AA]          &  [sec]      &        \\
\hline
\noalign{\hrule height 0.4pt}
A & La Silla/ESO & 2008 May  & 54603.68 & 2.2 & FEROS   &  0.03 & 3700-8600 &  600 &  70-120 \\
B & CASLEO & 2009 April & 54941.70 & 2.1 & REOSC CD &  0.2 & 3700-6000 & 1200 & 20-50 \\
C & CASLEO & 2012 January & 55930.86 & 2.1 & REOSC CD & 0.2 & 3700-6000 & 1200 & 20-50 \\
D & CASLEO & 2012 April & 56047.66 & 2.1& REOSC SD& 1.6 & 5900-7400 & 700 & 150-200 	\\
E & CASLEO & 2012 June & 56086.00 & 2.1& Boller\&Chivens & 2.7 &   3500-5000 & 150 & 160-250  \\
\hline
\end{tabular}
\end{table*}
\end{center}
}

\subsection{Photometric data}

The photometric data were taken from two sources: the Epoch Photometry Annex of Hipparcos Astrometry 
Mission (Perryman \& ESA, 1997) and The All Sky Automated Survey (ASAS) (Pojmanski, 1997).

The Hipparcos data consist of 214 observations made between December 1989  and January 1993. 
The measurements are in the interval $7.31 < H_p < 7.48$, while the errors of the individual 
measurements ranges from 0.007 to 0.020 mag. 
According to the criteria adopted when Hipparcos data were analyzed, this star was classified as an 
``unsolved variable''\footnote{cf. The Hipparcos and Tycho Catalogues, Vol. 1, Introduction and 
Guide to the Data, Sec. 1.3 App. 2.}. 
Following the Hipparcos calibration for a star with colour index $V-I = 0.070$, as in this case, 
the relation $H_p - V_J = 0.025$ should hold, where $V_J$ is the magnitude in the $V$ filter of the 
Johnson photometric system.

The ASAS data include 1030 magnitude determinations obtained between November 2000 and 
December 2009.
ASAS provides aperture photometry with several apertures. For this study, the aperture 
labeled as MAG~1 in the dataset was selected because larger apertures could 
be contaminated by a neighbor star. 
The errors of individual measurements 
are up to 0.120 mag, with a typical value err$_{ASAS} = 0.040$ mag. 
Since err$_{ASAS} > (H_p - V_J)$, no zero-point correction was applied to compare both datasets.

\section{Results}
\subsection{Spectral Analysis}

The FEROS spectrum (labelled A in Table~1) shows typical Be features as the Balmer 
lines H$\gamma$, H$\beta$ and H$\alpha$ in emission (see Fig~2). In it, is possible to see at least 1 additional
emission line, Fe~{\sc i} $\lambda$4328  bluer than H$\gamma$ and 2 additional emission lines, 
Fe~{\sc i} $\lambda$4849 and Fe~{\sc i} $\lambda$4875 of H$\beta$.
It also presents others metallic lines, i.e. O~{\sc i\sc i}, Fe~{\sc i\sc i}, 
Cr~{\sc i\sc i}, Ti~{\sc i\sc i\sc i}, C~{\sc i\sc i\sc i},
etc, as weak emissions, some of them shown in Fig~1.
All of these emission lines could be originated in a disk.
The double structure observed in H$\alpha$ emission lines is thought as originated by the Be star 
seen with mid inclination angle. 
The rotation of the disk plus possible absorption 
(``reversal" of the emission) against the edge of the star or disk cause such profiles. 
Other features as He~{\sc i}, $\lambda$4009, 4026, 4144, 4387, 4471, 4921, 5875, 6678 and 7065, 
Si~{\sc i\sc i} $\lambda\lambda$4128-30 and Mg~{\sc i\sc i} $\lambda$4481 are present in absorption. 
We interpreted that these lines are originated in the stellar photosphere. 
This composite spectrum indicates that HD\,112999's disk was thick and extended during 2008 May. 

\IBVSfig{7cm}{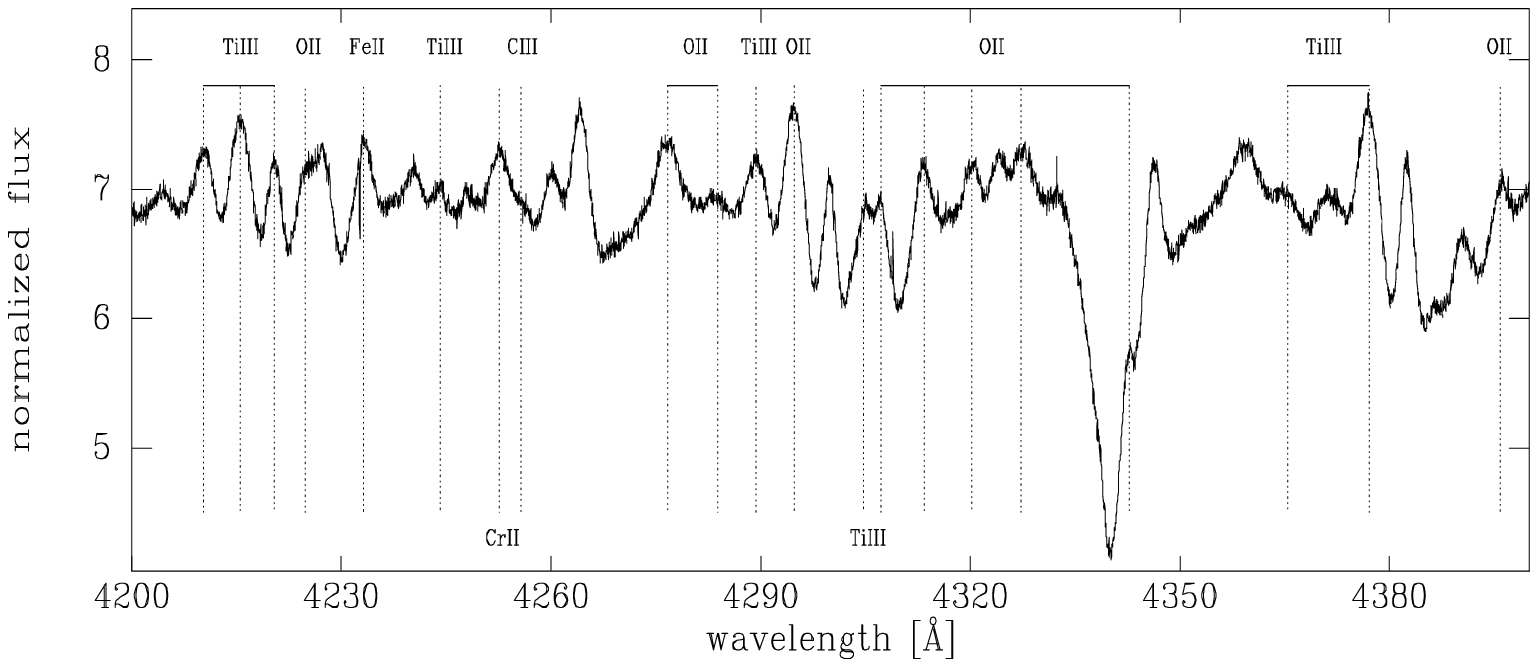}{Some metallic emission lines in the FEROS spectrum.}
\IBVSfigKey{FEROS1.eps}{HD112999}{Be star}

Our B spectrum, obtained about one year later than the A one, 
shows the emission contribution very weak in the H$\beta$ profile and almost marginal 
in the H$\gamma$ line. The metallic lines (seen as weak emissions in A) are not detected in this spectrum.
We think that this is not due to the B spectrum has poorer S/N ratio than the A one. The absorption lines, 
identified as arising in the stellar photosphere in the A spectrum, are also present here.
  
Three years later, the C spectrum presents the Balmer lines H$\gamma$ and H$\beta$ as pure absorption, 
i.e. there are no traces of the emission contribution (see Fig~2).

The D spectrum was obtained by one of us (WAW) to see the profile of the H$\alpha$ line, as none of B and C spectra
included it in their spectral ranges. 
Thus, we proved that H$\alpha$ is a conspicuous absorption line in the spectrum taken during
April 2012 (see Fig~2). 

The E spectrum is the one we have been able to obtain more recently (shown in Fig~3).
 It shows the same absorption spectral lines than the other spectra, plus a very narrow absorption line 
 at 4233 \AA , which we identified as Fe~{\sc ii}.
We consider this line was originated in a shell because of its FWHM is $\sim2.5$ \AA\, compared to the 
FWHM $\sim8$ \AA\, of the photospheric He~{\sc i} absorption lines.    
This asumption is in good agreement with the already proved by Arias et al. (2006) that the forming region of 
Fe~{\sc ii} lines is located close to the central star. 

We classified the Be-type spectra A, B, and E as B2$-$3~V, following the criteria of 
Walborn \& Fitzpatrick (1990),  
i.e. the ratio between Si~{\sc iii} $\lambda$4552 and Si~{\sc iv} $\lambda$4089, the
declining strengths of the C~{\sc iii}+O~{\sc ii} blends at $\lambda\lambda$4070 and 4650, 
and the weakness of all the Si~{\sc ii-iii} features.  
From the spectral type B2 to B3, Si~{\sc ii} $\lambda\lambda$4128-30, 
C~{\sc ii} $\lambda$4267 and Mg~{\sc ii} $\lambda$4481 increase in prominence. 
The very large He~{\sc i} $\lambda\lambda$4144/4121 ratio is a characteristic of the B2~V spectra.
The principal luminosity criterion is Si~{\sc iii} $\lambda$4552/ He~{\sc i} $\lambda$4387.

An important parameter to know in this class of star is the projected rotational 
velocity ($v \sin i$). To estimate this parameter, 
we have used the spectrum of the star $\tau$ Sco. It was convolved with rotation 
line profiles calculated for different projected rotational velocities and the 
\emph{FWHM} of the absorption line He~{\sc i} $\lambda$5015 was measured for each velocity.
Then, a linear relation between \emph{FWHM} and the $v \sin i$ was fitted. This empirical 
relation was used to translate the 
\emph{FWHM} of the line in the spectrum of HD~112999 into a $v \sin i$ estimation. 
In this way we obtained $v \sin i = 170 \pm 40$ km s$^{-1}$, being the error so 
large because of the low S/N ratio of the spectrum and the weakness of the line.

On the other hand, the emission of the star could be the result of disk instabilities 
caused by the spun up by binarity mass transfer (McSwain et al., 2009); thus, 
the B star becomes a Be star of rapid rotation. 
We measured the radial velocities in the three high-resolution spectra (A, B, and C). 
For this, we fit Gaussian profiles to selected spectral lines, 
i.e. He~{\sc i} $\lambda\lambda$4026, 4471, and 5876, and Mg~II $\lambda$4481  
as well as the interstellar lines of Na {\sc i}. On comparing the values of Na {\sc i}, 
no systematic difference is obtained among the spectrographs 
used (within an error of 3 km s$^{-1}$). However, the He~{\sc i} and Mg~{\sc ii}  $\lambda$4481 
absorption lines present variations of more than five times the expected error. 
Although, this result is not enough to confirm the binary nature of HD\,112999.
We provide the radial velocity measurements in Table 2 for future reference.

\begin{table}
\begin{center}
\caption{Radial velocities measured in the high resolution spectra}
\label{tab02}
\begin{tabular}{lccc}
\hline
Line & run A & run B & run C \\
 & [km s$^{-1}$] & [km s$^{-1}$] & [km s$^{-1}$] \\
\hline
He~I 4026 & -20 &   5 &     7  \\
He~I 4471 &  -35 & -15 & -31 \\
Mg~II  4481 & -92 & -5 & -2 \\
He~I 5876 &  -20 & 36 & -19 \\
Na~I 5890-96 & 4.5  & 5.5  & 2.5 \\
\hline
\end{tabular}
\end{center}
\end{table}

\IBVSfig{12cm}{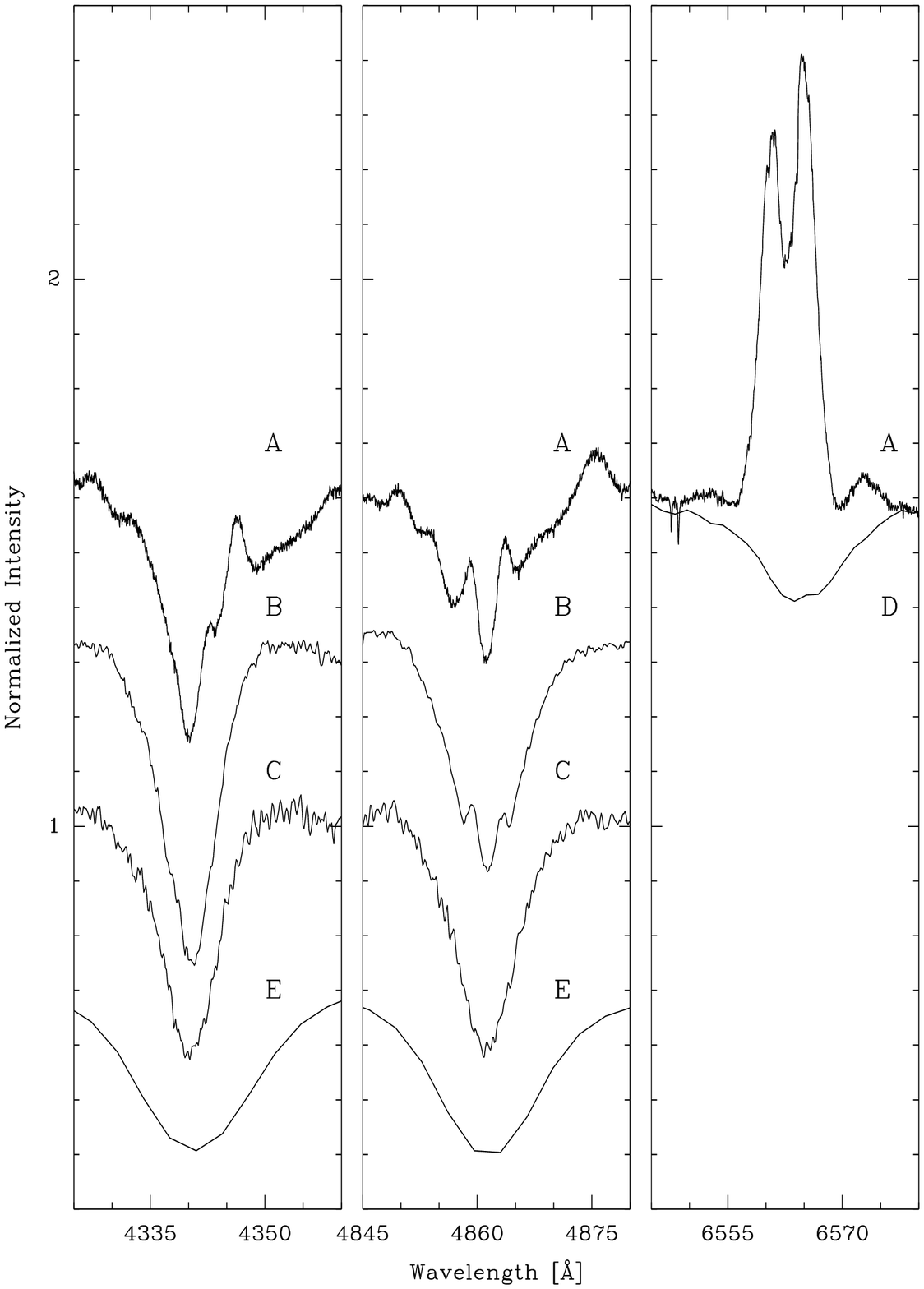}{Variations of H$\gamma$, H$\beta$ and H$\alpha$ lines from spectra obtained in 
different runs (see Table~1).
Other emission lines are also present. We identified them as 
Fe~{\sc i} $\lambda$ 4328 (in the H$\gamma$ region), 
Fe~{\sc i} $\lambda$ 4849 and $\lambda$ 4875 (H$\beta$ region), and
Fe~{\sc i} $\lambda$ 6574 (H$\alpha$ region).}
\IBVSfigKey{sgi27404.eps}{HD112999}{Be star}

\subsubsection{BCD fundamental parameters}

\IBVSfig{9cm}{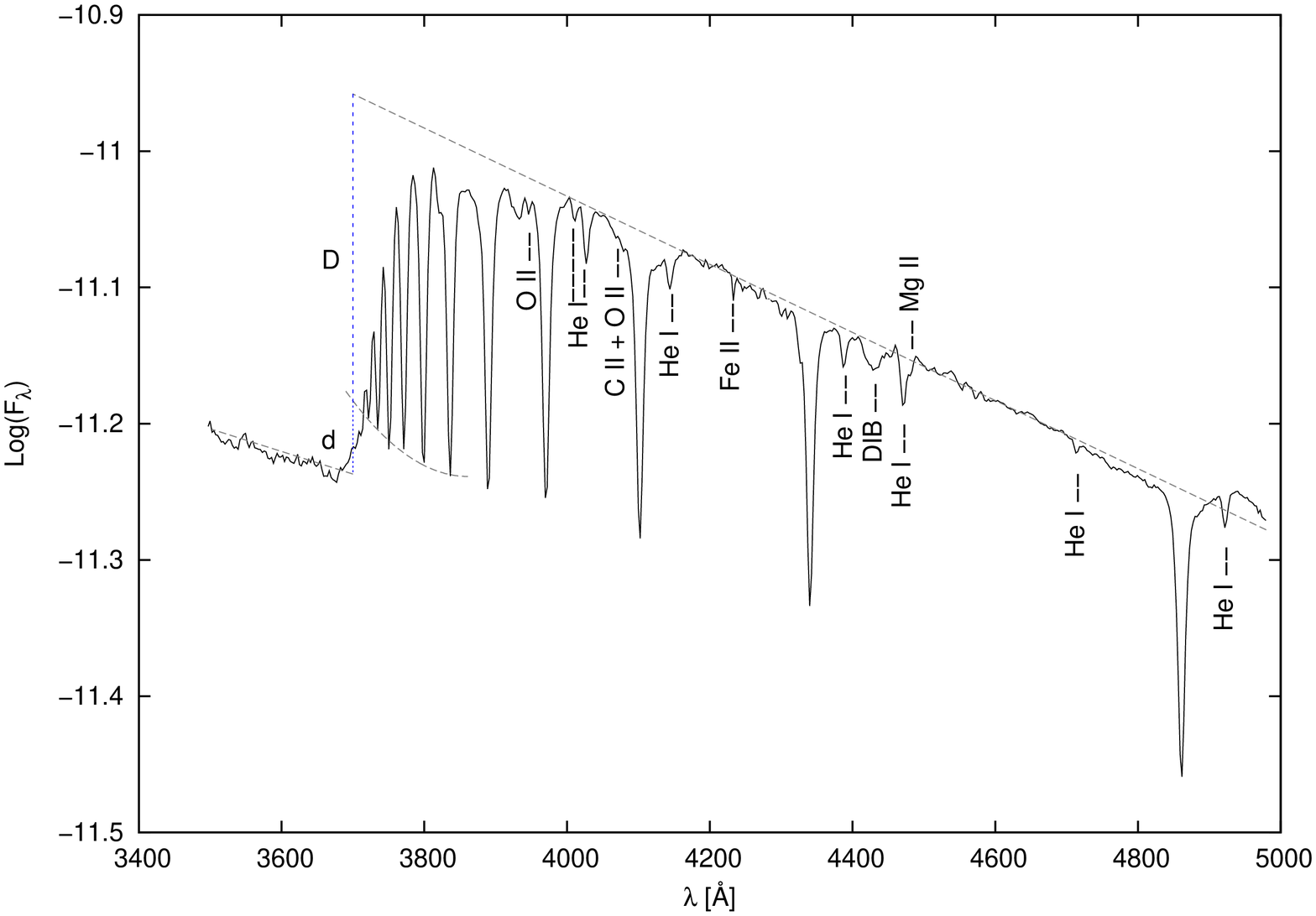}{Balmer discontinuity: D is the first Balmer jump for normal B
star, d is the second discontinuity, in absorption, which indicates the
presence of the circumstellar envelope.}
\IBVSfigKey{HD112999.BCD.eps}{HD112999}{Be star}

We derived some physical properties from the E spectrum of HD\,112999 using the BCD 
(Barbier-Chalonge-Divan) spectrophotometric system (Chalonge \& Divan, 1973, 1977). 
A detailed description of this method is presented in Zorec et al. (2009, Appendix A) and further 
applications to Be stars were done by Zorec et al. (2005) and Aidelman et al. (2012). 
This method is ideal for studying peculiar stars, and in particular Be stars, due to the fact 
that the parameters defined by the method
are not affected by interstellar extinction and absorption/emission from the circumstellar 
envelope (Zorec \& Briot, 1991).
Then, we determined a B3 spectral type (with one subtype error) of
luminosity class III, an effective temperature $T_{eff} = 17\,500 \pm 1\,000$~K, 
a surface gravity $log~g = 3.3 \pm$ 0.5, 
an absolute magnitude $M_V = -2.5 \pm$ 0.5 mag, 
and a color excess E$_{B-V} = 0.11 \pm$ 0.07 mag. 

The presence of the second Balmer discontinuity, in the E spectrum, indicates that
a circumstellar envelope is still present (Divan, 1979). Fig~3 
shows this spectrum in the Balmer discontinuity region, the first jump,  D, corresponds to a normal 
B star, while the second one,  d, is originated by the envelope.

We estimated the distance to HD~112999, using its extreme V magnitudes, i.e. 
7.3 $\leq$ V $\leq$ 7.5 (see Fig.~4), corrected by colour excess 
E$_{B-V} = 0.11 \pm$ 0.07 mag and considering the standard selective absorption coefficient, 
$R_{\mathrm v}$ = 3.1 (Schultz \& Wiemer, 1975). 
We obtained a distance modulus of 9.5 $\pm$ 0.5 mag  (790 $\pm$ 180 pc) 
and 9.7 $\pm$ 0.5 mag (870 $\pm$ 200 pc), respectively. 
 These estimated distance values with its respective errors are slightly 
larger than the Hipparcos distance (654 pc).  
Then, HD\,112999 results closer than the Cen\,OB1 association, $V_0 - M_v$ = 12 mag 
(2500 pc) (Corti et al., 2012).

\subsection{Light curve analysis}

\IBVSfig{9cm}{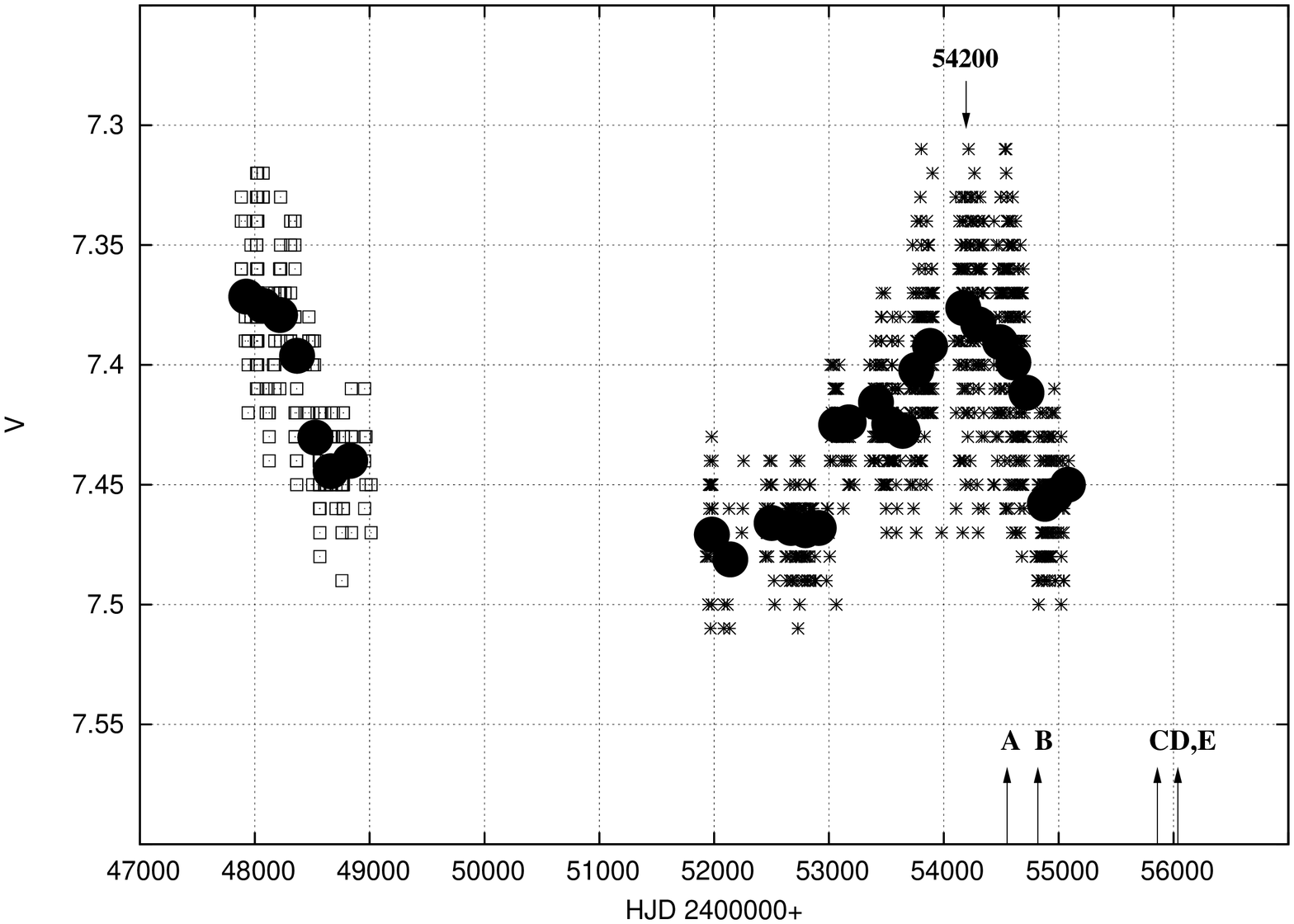}{Light curve of HD\,112999. 
Boxes and asterisks are Hipparcos and ASAS data, respectively. 
Filled circles depict the average magnitude in about 114 days. 
The capital letters (see Table~1) show the almost co$-$evality between the 
photometric and spectroscopic data explained in section 3.2.}
\IBVSfigKey{curvadeluz121109.eps}{HD112999}{Be star}

The collected photometric data were analyzed together and the resulting light curve is shown in 
Fig.~4, HD\,112999 presents two kinds of variabilities:  
a short-term (1.302 days; Hubert \& Floquet, 1998) and a long-term one ($\sim$ years). 
The former, smaller than 0.1 mag, is present in the Hipparcos data (see Fig.~5), and
it is the reason why this star was included in the General Catalogue of Variable Stars  
(Samus et al., 2009) and classified as BE\footnote{In the cases when a Be variable cannot be readily described 
as a Gamma Cassiopeiae variable (GCAS) star, (Samus et al. 2009) give simply BE for the type of variability}.

\IBVSfig{9cm}{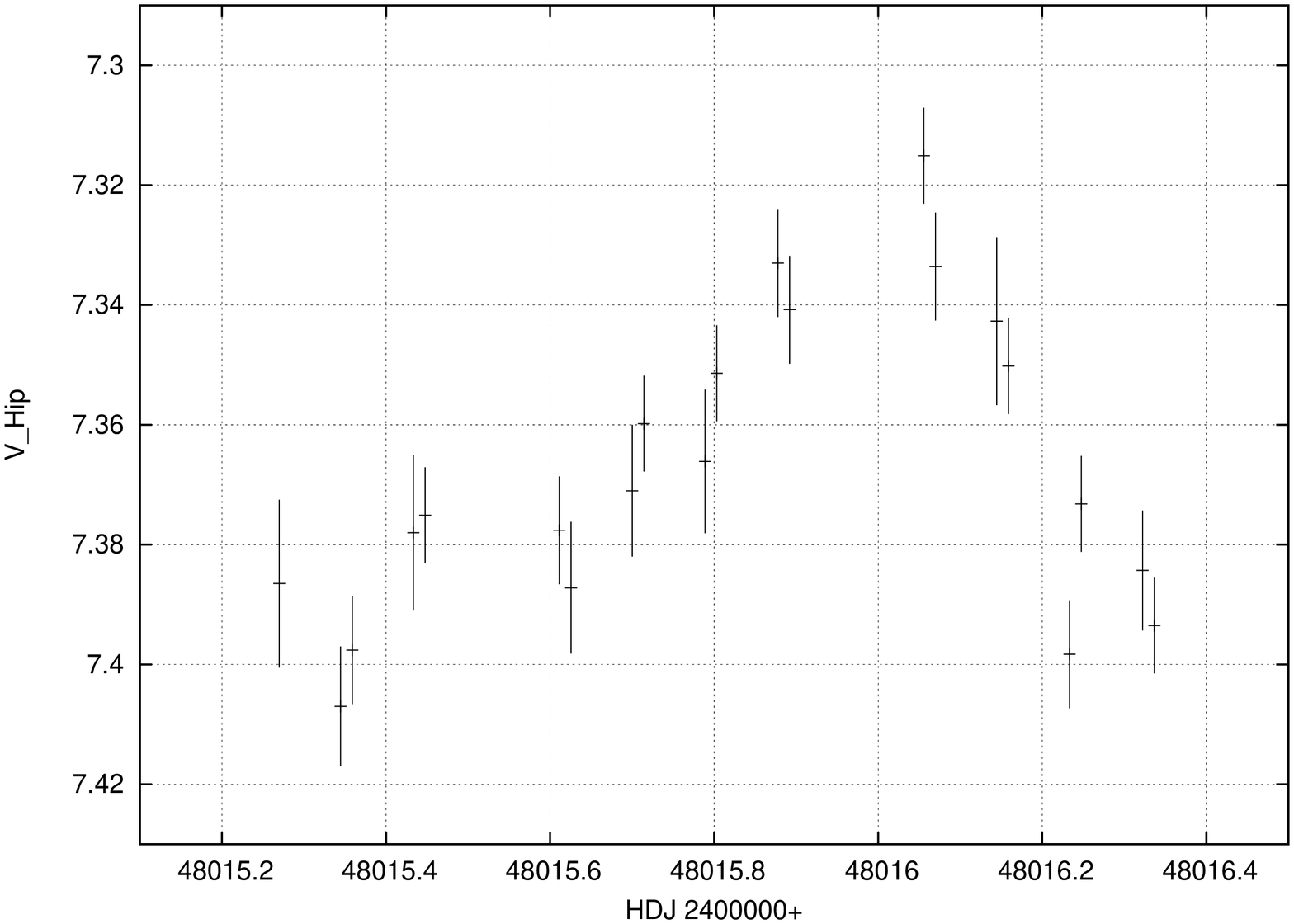}{Hipparcos data showing variability smaller than 0.1 mag in a period $\sim$ day.}
\IBVSfigKey{HD112999.photo.eps}{HD112999}{Be star}

The long-term variation can be better observed in the ASAS data.  
To enhance and show it more clearly, we represent unweighted averages  
(calculated within intervals of $\sim$114 days) with filled circles. 
In Fig.~4, we identified the light maximum reached on $\sim$ HJD 2\,454\,200 (ASAS data).

\section {Discussion}

The collected spectra show important variations, very noticeable in the Balmer lines, i.e.  
their emission profiles decrease in intensity from May 2008 to June 2012 (A to E). 
This kind of variability resembles the one already detected in other Be stars like MG\,31 and MG\,119 
in NGC\,3766 (McSwain et al., 2008); Pleione and $\kappa$~Dra 
Hubert \& Floquet, 1998) and 66~Ophiuchi (Floquet et al., 2002). 

The relation between the photometric and spectroscopic data is shown in Fig.~4.
The capital letters are the same defined in Table~1. 
The light curve and the spectra obtained in A and B runs, are well correlated in the sense that as the 
star fades, 
the Balmer emissions decrease their intensities (Fig.~2). This behavior is
related to disk dissipation. The lack of photometric data after 2009 August prevents us to determine 
if the star reached the minimum between B and C runs or it will be reached later.

The singular behavior observed in HD\,112999 led us to search in the existing literature in order to 
become familiar with the information already known about this star. Thus, we learned that the
unique Be spectrum of HD\,112999 was obtained in the period 1967$-$1969 by 
Bidelman \& MacConnell (1973)\footnote{This observation was the reason to 
be included in the Catalogue of Be Stars (Jaschek \& Egret, 1982)}.
Unlikely, they did not provide a precise date of observation, which enables us to relate it with the 
date of the FEROS spectrum.
As said in Section~1, other authors have reported different spectral classifications for HD\,112999.
We adopted the B2$-$3~V spectral$-$type for HD\,112999, determined following the criteria of 
Walborn \& Fitzpatrick (1990).

\section{Summary}

The analysis of the spectroscopic and photometric data of HD\,112999 allowed us to identify 
a long$-$term variability in its $V$ light curve and the change of the profiles of the Balmer
lines from emission to absorption. The declining on V$-$magnitudes between 2007 and 2009 
is related with the disappearance of emission lines, very noticeable in some Balmer lines.
We propose that the reported changes are originated in the loss of the disk.
Photometric data were not sufficient to search for long-term periodicities, thus we encourage other
researchers to monitor this star in the future. 
We classified our spectra as B2$-$3, the luminosity class should be clarified with further 
observations. 

We took advantage of the observations obtained on different observing programs to measure some
important parameters as the projected rotational velocity, and the distance modulus.  
Also, our last (E) spectrum, obtained on 2012 June, was analyzed via the BCD method and noted 
that a circumstellar envelope is still present. Hence, further spectra should be obtained to 
determine if HD\,112999 will reach a normal B$-$type. \\

\noindent ACKNOWLEDGMENTS:\\
This research has made use of the SIMBAD database, operated at CDS, Strasbourg, France, and also
the WEBDA database, operated at the Institute for Astronomy of the University of Vienna.
This work was partially supported  by Consejo Nacional de Investigaciones 
Cient\'{\i}ficas y T\'ecnicas {\rm CONICET} under projects {\rm PIP} 
01299 and 0523 and by Universidad Nacional de La Plata ({\rm UNLP})
under project G091. We are very grateful to Dra. Granada Anah\'{\i} and
Dra. Cidale Lydia for their interesting and valuable comments. We thank Mrs Roxana Kotton 
and the English Department of FCAyG (UNLP) for a careful reading of the manuscript.

\references

Aidelman, Y., Cidale, L.~S., Zorec, J., Arias, M.~L., 2012, {\it A\&A}, {\bf 544}, A64

Arias, M.~L., Zorec, J., Cidale, L.~S. et al., 2006, {\it A\&A}, {\bf 460}, 821 

Bidelman, W.~P., MacConnell, D.~J., 1973, {\it AJ}, {\bf 78}, 687

Cannon, A.~J., Pickering, E.~C, 1920, {\it Annals of Harvard College Observatory}, {\bf 95}, 1

Chalonge, D., Divan, L, 1973, {\it A\&A}, {\bf 23}, 69

Chalonge, D., Divan, L, 1977, {\it A\&A}, {\bf 55}, 117

Corti, M.~A., Arnal, E.~M., Orellana, R.~B, 2012, {\it A\&A}, {\bf 546}, A62

Divan, L, 1979, {\it in Ricerche Astronomiche}, Vol. {\bf 9}, IAU Colloq. 47: Spectral Classification of the Future, 
ed. M.~F.~McCarthy, A.~G.~D.~Philip, \& G.~V.~Coyne, 247-256

 Floquet, M, Neiner, C,  Janot-Pacheco, E, Hubert, A.~M, et al., 2002, {\it A\&A}, {\bf 394}, 137

Garrison, R.~F., Hiltner, W.~A., Schild, R.~E., 1977, {\it ApJS}, {\bf 35}, 111

Houk, N., Cowley, A.~P, 1975, {\it University of Michigan Catalogue of two-dimensional spectral types for the HD stars.} Volume {\bf I}. Declinations -90 to -53

Huat, A.~L., Hubert, A.~M., Baudin, F., Floquet, M, Neiner, C, et al., 2009, {\it A\&A}, {\bf 506}, 95

Hubert, A.~M., Floquet, M, 1998, {\it A\&A}, {\bf 335}, 565

 Jaschek, M., Slettebak, A., Jaschek, C., 1981, {\it BeSN}, {\bf 4}, 9

Jaschek, M., Egret, D., 1982, {\it IAU Symposium}, Vol. {\bf 98}, Be Stars, ed. Jaschek, M., Groth, H.-G., 261

McSwain, M.~V., Huang, W., Gies, D.~R., 2009, {\it ApJ}, {\bf 700}, 1216

McSwain, M.~V., Huang, W., Gies, D.~R., Grundstrom, E.~D., Townsend, R.~H.~D., 2008, {\it ApJ}, {\bf 672}, 590

 Neiner, C., de Batz, B, Cochard, F, Floquet, M., Mekkas, A, Desnoux, V, 2011, {\it AJ}, {\bf 142}, 149

Neiner, C., Grunhut, J.~H., Petit, V., ud-Doula, A., Wade, G.~A., et al., 2012, {\it MNRAS}, {\bf 426}, 2738.

Perryman, M.~A.~C., ESA, eds. 1997, {\it ESA Special Publication}, Vol. {\bf 1200}, 
The HIPPARCOS and TYCHO catalogues. 
Astrometric and photometric star catalogues derived from the ESA HIPPARCOS Space Astrometry Mission

Pojmanski, G., 1997, {\it Acta Astron.} {\bf 47}, 467

Porter, J.~M., Rivinius, T, 2003, {\it PASP}, {\bf 115}, 1153

Samus, N.~N., Durlevich, O.~V., et al., 2009, {\it VizieR Online Data Catalog}, {\bf 1}, 2025

Schultz, G.~V., Wiemer, W., 1975, {\it A\&A}, {\bf 43}, 133

Slettebak, A., Collins, II, G.~W., Parkinson, T.~D., Boyce, P.~B., White, N.~M., 1975, {\it ApJSS}, {\bf 29}, 137

Walborn, N., Fitzpatrick, E., 1990, {\it PASP}, {\bf 102}, 379

Zorec, J., Briot, D., 1991, {\it A\&A}, {\bf 245}, 150

Zorec, J., Cidale, L., Arias, M.~L. et al., 2009, {\it A\&A}, {\bf 501}, 297

Zorec, J., Fr\'emat, Y., Cidale, L., 2005, {\it A\&A}, {\bf 441}, 235

\endreferences

\end{document}